\documentclass[twocolumn,showpacs,preprintnumbers,amsmath,amssymb,showkeys]{revtex4}
\usepackage{graphicx}
\usepackage{bm}
 
\newcommand{\Xs}{X_{\rm S}}
\newcommand{\vc}{v_{\rm c}}

\DeclareMathOperator{\sech}{sech}
\DeclareMathOperator{\trig}{trig}

\begin{document}

\title{Chaotic Scattering and the $n$-bounce Resonance in Solitary Wave Interactions}

\author{Roy H. Goodman}
 \email{goodman@njit.edu}
\affiliation{%
Department of Mathematical Sciences, New Jersey Institute of
  Technology, Newark, NJ 07102}%

\author{Richard Haberman}
 \email{rhaberma@smu.edu}
\affiliation{Department of Mathematics, Southern Methodist University, Dallas, 
TX 75275}%

\date{\today}%

\begin{abstract}
We present a new and complete analysis of the $n$-bounce resonance and chaotic scattering in solitary wave collisions. In these phenomena, the speed at which a wave exits a collision depends in a complicated fractal way on its input speed.  We present a new asymptotic analysis of collective-coordinate ODEs, reduced models that reproduce the dynamics of these systems. We  reduce the ODEs to discrete-time iterated separatrix maps and obtain new quantitative results unraveling the fractal structure of the scattering behavior. These phenomena have been observed repeatedly in many solitary-wave systems over 25 years.
\end{abstract}

\pacs{05.45Yv, 
05.45Gg 
}
\keywords{solitons, chaos, solitary wave collisions}
\maketitle

Solitary waves---localized disturbances that travel with unchanging shape and velocity---are ubiquitous in physical science, and are seen, for example, in fluid mechanics, optics, solid-state electronics, and even quantum field theory.  A natural question is what happens when the wave hits an obstacle or two such waves collide.  

In dissipative systems such as electrical signal propagation in nerve fibers or reaction-diffusion systems, two interacting waves generally merge into a single larger wave.  In completely integrable, or soliton, equations, by contrast, interacting solitary waves emerge from a collision intact and with their original speeds, but a slight shift in their position, which is well-understood through the theory of inverse scattering.  

\begin{figure}
\begin{center}
\includegraphics[width=3in]{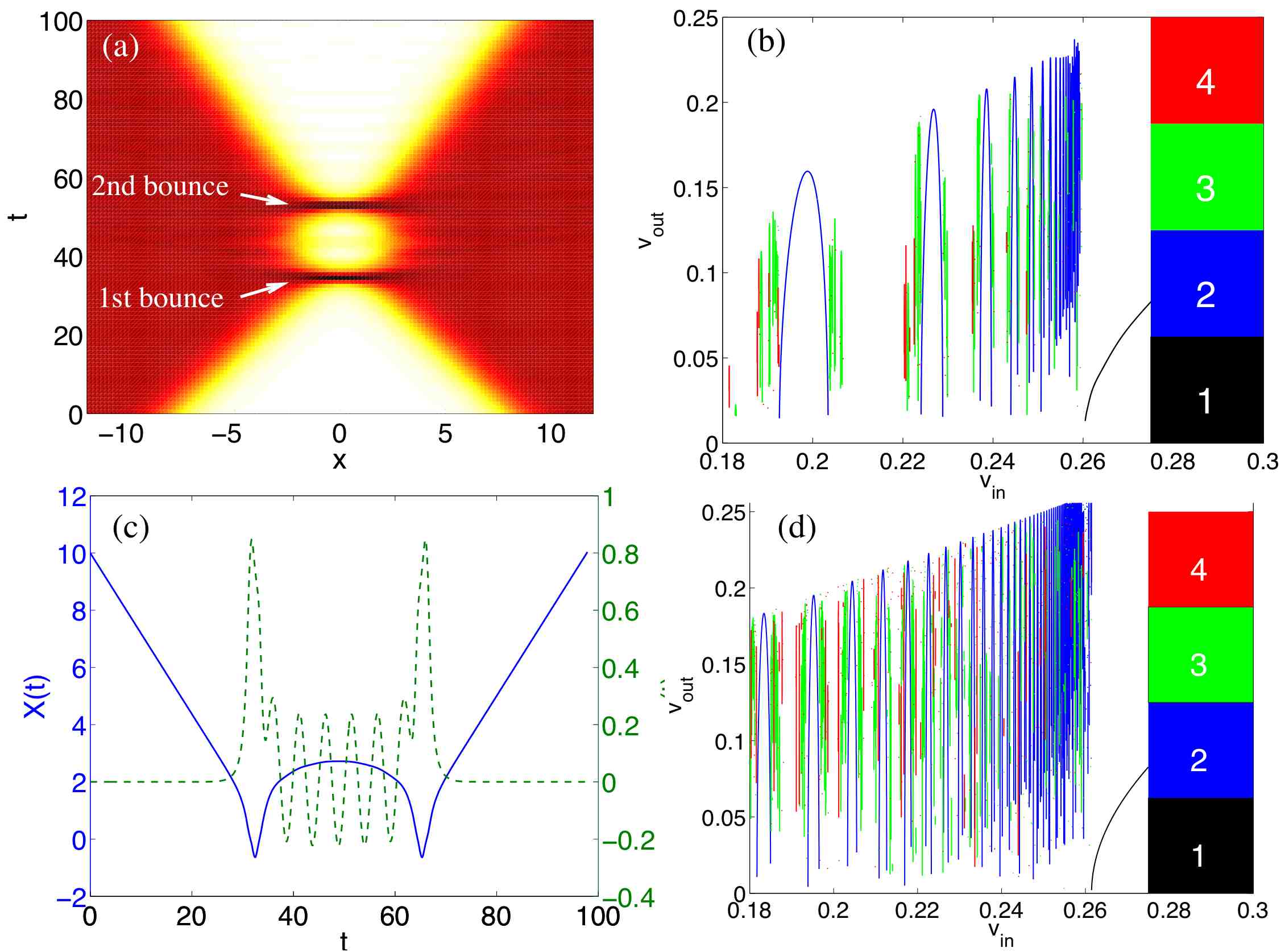}
\caption{Color online: (a) A `2-bounce' solution to PDE~\eqref{eq:phi4}. (b)  $v_{\rm in}$ vs.\ $v_{\rm out}$ for kink-antikink collisions to~\eqref{eq:phi4} showing chaotic scattering. 
(c) Two-bounce solution to ODE model for~\eqref{eq:phi4} ($X(t)$ solid $A(t)$  dashed.) (d) $v_{\rm in}$ vs.\ $v_{\rm out}$ for the ODE model, annotated as in (b).}
\label{fig:2bounce}
\end{center}
\end{figure}

Collisions in dispersive wave systems that are neither dissipative nor completely integrable may produce a much wider range of behaviors.  We focus on one, the 2-bounce, or, more generally $n$-bounce phenomenon. Two counterpropagating waves with sufficient relative initial speed (or one wave incident on a localized defect) will pass by or reflect off each other with little interaction, while for most initial speeds below some critical velocity $\vc$ they will become trapped, forming a localized bound state.  At certain velocities below $\vc$, the waves become trapped, begin to move apart, and come together a second time before finally moving apart for good---the so-called 2-bounce solutions.    In addition to the 2-bounce resonant solutions, one often finds 3-, 4-, or, more generally, $n$-bounce solutions.  Figure~\ref{fig:2bounce}a shows a 2-bounce resonant solution to~\eqref{eq:phi4}, and figure~\ref{fig:2bounce}b shows the sensitive dependence of the final speed on the initial speed, with the number of `bounces' indicated by color. The initial conditions leading to these behaviors are interleaved in a manner often described as fractal.  This was first seen in kink-antikink collisions in the mid 1970's (see~\cite{CP} and references therein), subsequently found in models from astrophysics~\cite{AOM:91}, optical fiber communications~\cite{TY} and perhaps most recently in 2007 in collisions between topological solitons arising in quantum field theory\cite{PieZak:07}.  Figure~\ref{fig:2bounce}c shows a 2-bounce resonant solution of the model ODEs for system~\eqref{eq:phi4} (discussed below), and figure~\ref{fig:2bounce}d shows that the ODE model reproduces the fractal interaction structure of the PDE, if not the exact structure. 

We analyze these phenomena through systematic asymptotics applied to `collective coordinate' models, low-dimensional model systems of ordinary differential equations (ODEs) derived from a variational principle that reproduce the dynamics in numerical simulations.  We construct, using  Melnikov-integrals and formal matching procedures, approximate $n$-bounce resonant solutions to the ODEs and derive an iterated map that explains the fractal structure\cite{GooHab,GooHab:05}.  

Previous studies treat the results of numerical simulations (both ODE and PDE) as experimental data and have remarkable success analyzing these results using a combination of numerical simulation and ad-hoc calculations.  They derive approximate resonant velocities using least-squares fitting of numerical data.  By contrast, we obtain formulae dependent only on the equations' parameters, and not on any empirical constants.

The phenomenon was first observed in kink-antikink collisions in nonlinear wave equations by Campbell, Peyrard et al.~\cite{CP}, including the  $\phi^4$ equation,
\begin{equation}
u_{tt} - u_{xx} + u - u^3 = 0.
\label{eq:phi4}
\end{equation}
Figure~\ref{fig:2bounce}b is a new computation that reproduces one from their first paper.  What was in the early 1980's a very difficult and time-consuming computation we reproduced in a short time on a PC, with improved detail showing narrower $n$-bounce windows between the primary 2-bounce windows~\cite{GooHab:05}.
Fei, Kivshar, and V\'azquez subsequently observed 2-bounce solutions in collisions of kinks with Dirac delta potentials in the sine-Gordon and $\phi^4$ equations~\cite{FKV}, 
\begin{align}
u_{tt} - u_{xx} +(1-\epsilon \delta(x))\sin u&=0 \mbox{ and} \label{eq:sG}\\
u_{tt} - u_{xx} +(1- \epsilon \delta(x))(u-u^3&)=0. \label{eq:phi4def}
\end{align}
Tan and Yang saw it in collisions between orthogonally-polarized solitons in birefringent optical fibers~\cite{TY}, described by coupled nonlinear Schr\"odinger equations:
\begin{equation}
i \partial_t u_i + \partial_x^2 u_i + (|u_i|^2 + \epsilon |u_{1-i}|^2) u_i = 0, \, i =0,1.
\label{eq:nls2}
\end{equation}

What all these (non-integrable) dispersive wave equations have in common is a second mode which can draw energy from the propagating wave. When the solitary wave is taken to model a pseudo-particle, this corresponds to an internal oscillatory degree-of freedom.
This transfer creates an effective energy barrier, preventing slow waves from escaping the collision location.  

\begin{figure}[b]
\begin{center}
\includegraphics[width=3.25in]{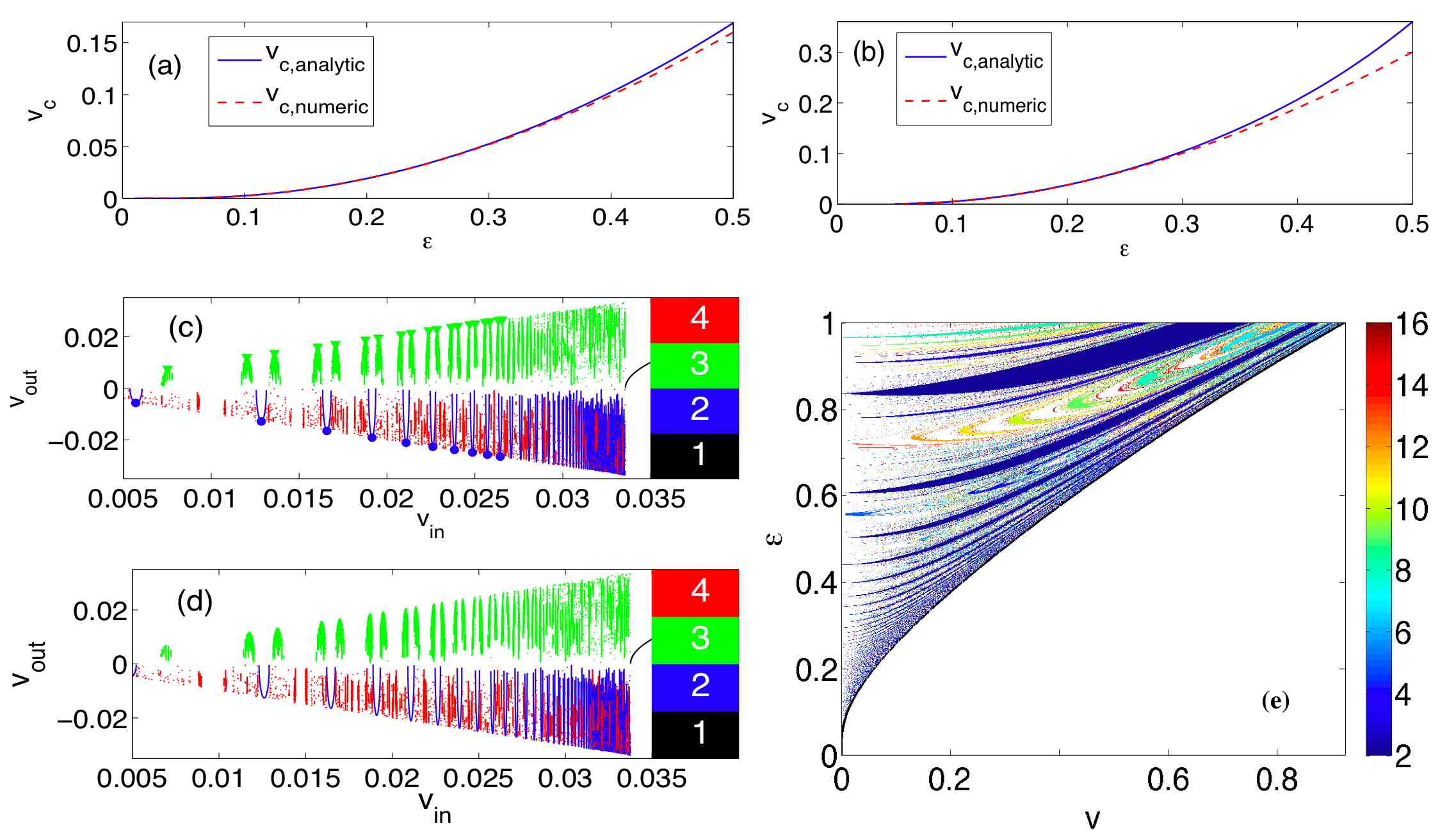}
\caption{(Color Online) (a): The critical velocity $\vc$ for  model~\eqref{eq:sg_defs}. (b): As (a), for model~\eqref{eq:phi4_model}. (c): $v_{\rm in}$ vs. $v_{\rm out}$  for the discrete map with the number of bounces coded by color;  from~\eqref{eq:vn} $v_n^{(2)}$ are marked by $\bullet$ and $v_{n\pm}^{(3)}$ by $\blacktriangledown$.  (d): As (c), for numerical integration of ODE~\eqref{eq:sg_defs}. (e): Number of bounces as a function of both $\epsilon$ and the initial velocity.}
\label{fig:sgmap}
\end{center}
\end{figure}

In the following paragraphs, we analyze the behavior shown in figures~\ref{fig:2bounce}c and~\ref{fig:2bounce}d.  The general  form of the ODE  model is given below in~\eqref{eq:general_ode}.  We provide the critical velocity for capture in~\eqref{eq:vc} and displayed in figure~\ref{fig:sgmap}a and~\ref{fig:sgmap}b.  The locations of the 2-bounce windows, and the narrower 3-bounce windows is given in equation~\eqref{eq:vn} as special solutions of an iterated map we define below.

Each system above is well-known to possess a variational form~\cite{Mal:02}: their solutions minimize a Lagrangian
\begin{equation}
L(u,x,t) = \int_{t_1}^{t_2} \int_{-\infty}^{\infty} {\cal L}\bigl({u(x,t)\bigr)} \ dx \ dt.
\label{eq:lagrange}
\end{equation}
ODEs are derived by assuming the solution depends on a few time-dependent parameters $u(x,t) \approx u_{\rm ansatz}(X_1(t),\ldots,X_n(t))$, inserting this ansatz into integral~\eqref{eq:lagrange} and integrating out the $x$-dependence to obtain a finite-dimensional Lagrangian whose Euler-Lagrange equations describe the evolution of the parameters $\vec X(t)$.

For systems~\eqref{eq:phi4}-\eqref{eq:nls2}, the ansatz depends on a variable $X(t)$, parameterizing the distance between the two interacting solitary pulses in systems~\eqref{eq:phi4} and~\eqref{eq:nls2} and the pulse position in~\eqref{eq:sG} and~\eqref{eq:phi4def}, and a variable $A(t)$ measuring the amplitude of a second mode of oscillation---two such modes in the ODE model equations for system~\eqref{eq:phi4def}.  The ODE models take the general form (after some rescalings)
\begin{equation}
\label{eq:general_ode}
m \ddot X + U'(X) + F'(X) A  =0;\;
\ddot A + \omega^2 A + c F(X) =0,
\end{equation}
where $c$ or $\omega^{-1}$ is a small parameter, allowing the use of perturbation methods.
The ODE model for systems~\eqref{eq:phi4} and~\eqref{eq:phi4def} contain additional terms but can be treated using the methods described herein.  We refine our terms describing the ODE model: an $n$-bounce solution is one in which $X(t)$ escapes to infinity after $n$ interactions, and $n$-bounce resonance is such a solution for which, additionally $A(t) \to 0$ as $t \to \pm \infty$ so that $v_{\rm in} = v_{\rm out}$.

We consider two such models here.  After rescaling time, system~\eqref{eq:sG} is modeled by the equations
\begin{equation}
 U(X) = -2 \sech^2{X}; \;  F(X) = -2 \tanh{X}\sech{X};  \label{eq:sg_defs}
 \end{equation}
 where $m =4$, $ \omega^2=2/\epsilon-\epsilon/2$  and $c=\epsilon$.
The ODEs that model equation~\eqref{eq:phi4} are algebraically complex and are studied in~\cite{AOM:91,GooHab:05}, but their essential dynamics (determined by the topology of the phase-space) are captured by making $U(X)$ the Morse potential and choosing a simple $F(X)$ which vanishes at infinity: 
\begin{equation}
U(X)=e^{-2X}-e^{-X} \text{ and } F(X) = e^{-X}.
\label{eq:phi4_model}
\end{equation}
Equation~\eqref{eq:general_ode} conserves  $H= \frac{m}{2}\dot X^2 + U(X) + \frac{1}{2c}(\dot A^2 + \omega^2 A^2) + F(X) A$, where the first two terms describe energy in the propagating wave, the second two, energy in $A$ and the final one, the coupling energy.  

\begin{figure}[htb]
\begin{center}
\includegraphics[width=3in]{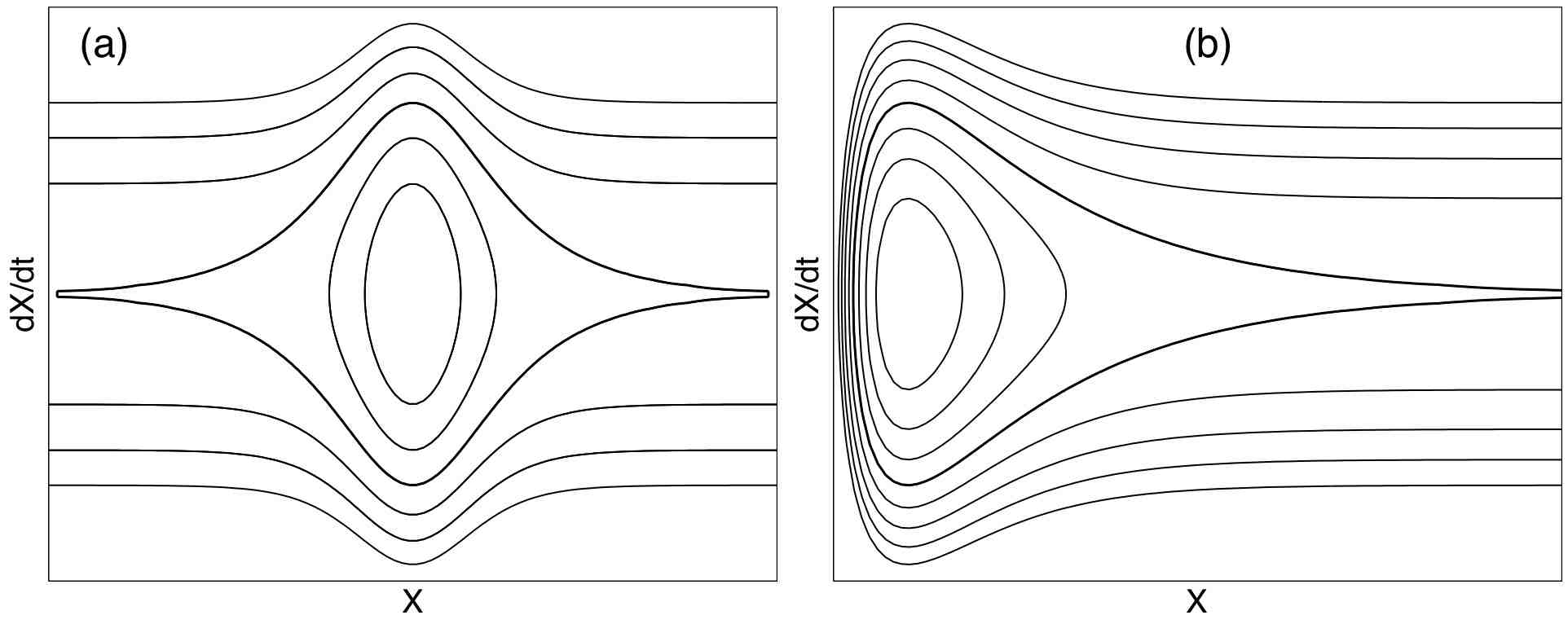}
\caption{The uncoupled phase portrait due to the potential $U(X)$ in (a) equations~\eqref{eq:sg_defs} and (b) equations~\eqref{eq:phi4_model}.}
\label{fig:phaseplane}
\end{center}
\end{figure}

The dynamics of $X(t)$ in~\eqref{eq:general_ode}, neglecting $A$, conserves an energy $E= \frac{m}{2}\dot X^2 + U(X)$, and the trajectories lie on level sets of $E$, figure~\ref{fig:phaseplane}, with $E>0$ along unbounded ($E<0$ along bounded) trajectories and separatrix orbits along which $E=0$.
In the first instance the phase plane has two heteroclinic orbits connecting degenerate (saddle-type) fixed points at $(X,\dot X) = (\pm \infty,0)$, while the second has one homoclinic to $(+\infty,0)$.

When $A(t)$ is allowed to vary, the level sets of $E$ cease to be invariant.  Define capture (escape) to be  a trajectory that crosses a separatrix from the region of unbounded trajectories to that of bounded trajectories (bounded to unbounded).  We construct approximate solutions via matched asymptotic approximations where `outer solutions' consist of expansions near the degenerate saddle points, which are connected  via `inner solutions,'  i.\ e.\ separatrix orbits.  An energy change calculated over each separatrix orbit is used to match together two consecutive outer approximations near infinity.

Over a full trajectory from one saddle-approach to the next, the total change in $E$ is the Melnikov integral~\cite{GH:83}
\begin{equation}\begin{split}
\Delta E &= \int_{-\infty}^\infty \frac{dE}{dt} dt =   \int_{-\infty}^\infty (m \ddot X + U'(X))\dot X dt\\
& =  -\int_{-\infty}^\infty A \frac{d}{dt}F(X(t)) dt   =  \int_{-\infty}^\infty F(X(t)) \frac{dA}{dt} dt 
\label{eq:DeltaE1}
\end{split}\end{equation}
assuming $F(X(t))\to 0$ as $|t| \to \infty$.
Equation~\eqref{eq:general_ode} may be solved for $A$ by variation of parameters and used to simplify~\eqref{eq:DeltaE1}.  Under the assumption that $A(t)\to 0$ as $t\to-\infty$, this is
 \begin{equation}
\label{eq:DeltaE}
\Delta E = -\frac{c}{2}  \left \lvert  \int_{-\infty}^\infty  F(X_{\rm S}(t))e^{i\omega t} dt \right \rvert ^2,
\end{equation}
 the Fourier transform of $F(\Xs(t))$ evaluated at $\omega$, the resonant frequency of $A(t)$. 
Here  $X(t)$ has been approximated by $X_{\rm S}(t)$, the solution
along the separatrix. The critical velocity, which solves $m \vc^2/2 = \Delta E$, is~\cite{GooHab, GooHab:05}:
\begin{equation} \label{eq:vc}
\vc = \sqrt{\frac{c}{m}}  \left \lvert  \int_{-\infty}^\infty  F(\Xs(t))e^{i\omega t} dt \right \rvert .
\end{equation}
For system~\eqref{eq:sg_defs}, $\Xs = \sinh^{-1}(t-t_1)$ and $\vc = \pi \sqrt\epsilon e^{-\omega}. $
While for system~\eqref{eq:phi4_model}, $\Xs = \log {\left(1+\frac{(t-t_1)^2}{2m}\right)}$ and $\vc = \sqrt{2c}\pi e^{-\omega\sqrt{2m}}.$ These are shown in figure~\ref{fig:sgmap}a and b (with $c=\epsilon^2$, $m=1$ and $\omega=\epsilon^{-1/2}$ in~\eqref{eq:phi4_model}).

The time $t_1$ is  the `symmetry time' of the first inner solution, at which $\Xs=0$ in model~\eqref{eq:sg_defs} and $\frac{d}{dt} \Xs =0$ in model~\eqref{eq:phi4_model}.  If $E>0$ after the first interaction, the wave moves off toward infinity. If $E<0$ the wave turns around. The oscillator $A(t)$ has become excited and the variation of parameters formula shows that, for large $t$,
\begin{equation}
\begin{split}
A(t) \sim &-\frac{c}{\omega} \sin{\omega(t-t_1)} \int_{-\infty}^{\infty} F(\Xs(\tau)) \cos{\omega(\tau-t_1)} d\tau \\
&+\frac{c}{\omega} \cos{\omega(t-t_1)}\int_{-\infty}^{\infty} F(\Xs(\tau))
\sin{\omega(\tau-t_1)} d\tau.
\label{eq:A_asympt}
\end{split}\end{equation}
In model~\eqref{eq:sg_defs}, $F(\Xs(t))$ is odd about $t_1$, so the first integral vanishes identically and $A(t) \propto \cos{\omega(t-t_1)}.$
In model~\eqref{eq:phi4_model}, $F(\Xs(t))$ is even about $t_1$, so  $A(t) \propto \sin{\omega(t-t_1)}.$ 
The solution now alternates between two behaviors---negative energy `outer solutions' dominated by the degenerate fixed point at $\pm\infty$ and near-separatrix solutions with center time $t_i$, until at some step $n\ge2$, $E_n>0$ and the pulse escapes.

Equation~\eqref{eq:DeltaE} gives $\Delta E$ along the first near-separatrix solution.  At each subsequent interaction, at time $t_i$, a similar calculation is performed~\cite{GooHab,GooHab:05}, with new terms that arise because $A(t)$ no longer approaches zero as in backward time along the near-separatrix solution, but instead is asymptotically given by a sum of terms like~\eqref{eq:A_asympt}, one for each previous collision.  The energy level $E$ after the $n$th interaction depends, thus, not only on the initial energy, but on the sequence of times $t_1$ through $t_n$.  The time difference $t_j - t_{j-1}$, in turn,  is a function of the energy level $E_{j-1}$, since the period of this nonlinear oscillator depends on its energy, as we show below. 

This time-change can be calculated by the matching conditions between the near-separatrix solution centered at $t_{j-1}$ and the near-saddle-expansion (outer solution) immediately following, and then the near saddle expansion to the next near-separatrix solution.  

Under the assumption that as $X \to \infty$,  $U(X) \sim -\alpha^2 e^{-2 \beta X}/2,$
we examine the large-$X$ behavior of the first near-separatrix solution. A divergent integral for $t-t_1$ is regularized as:
\begin{equation}
t-t_1  =\sqrt{\frac{m}{2}} \int_{X_1}^X \frac{dY}{\sqrt{-U(Y)} }\\
\approx  \frac{\sqrt m}{\alpha}  e^{\beta X} +{\mathcal R},
\label{eq:tt1}
\end{equation}
where ${\mathcal R} = \frac{\sqrt m}{2} \int_{X_1}^\infty \left(  \frac{1}{\sqrt{-U(Y)}}  - \frac{\sqrt{2}}{\alpha} e^{\beta Y} \right) dY - \frac{\sqrt m}{\alpha\beta} e^{\beta X_1}$.
A calculation for the ensuing large-$X$-saddle (outer) approximate solution with energy $E_1<0$, assumed to reach its maximum $X=X^*$ at $t=t^*$, yields
\begin{equation}
\cos{ \left(\sqrt {\frac{-2E_1}{m}}\beta(t^*-t)\right)} = \frac{\sqrt{-2E_1}}{\alpha} e^{\beta X}.
\label{eq:tstar}
\end{equation}
Matching~\eqref{eq:tstar} with~\eqref{eq:tt1} yields, via a consistency condition, $t^*-t_1$.
The calculation for $t_2-t^*$ is identical and eliminating $t^*$ gives $t_2-t_1$.
This calculation and its conclusion are unchanged for each time interval $(t_{j-1},t_j)$ and energy $E_{j-1}$, yielding
\begin{equation}
\label{eq:dt}
t_j-t_{j-1} =  2{\cal R} + \sqrt{\frac{m}{-2E_{j-1}}}\frac{\pi}{\beta}.
\end{equation}

As $t-t_j \to \infty$ along the $j$th near-separatrix solution, 
\begin{equation}
A(t) \sim \sum_{k=1}^{j} \sigma_k \frac{2\sqrt{\epsilon}}{\omega}\trig {\omega(t-t_k)},
\label{eq:A_series}
\end{equation}
where $\sigma_k \trig x = (-1)^{k+1} \cos x $ in model~\eqref{eq:sg_defs}  and $\sigma_k\trig x =- \sin x $ in model~\eqref{eq:phi4_model}.  This implies
\begin{equation}
\label{eq:DeltaE_general}
E_j = E_{j-1} -\frac{ m \vc^2}{2}  \bigl( 1 + 2 \sum_{k=1}^{j-1} \sigma_{j-k} \cos{\omega(t_j-t_k)} \bigr),
\end{equation}
with initial energy $E_0 = \frac{m v_0^2}{2}$.  Equations~\eqref{eq:dt} and~\eqref{eq:DeltaE_general}, applied alternately, constitute the separatrix map. 

Summing equation~\eqref{eq:DeltaE_general}, we find 
\begin{equation}
\label{eq:njumps}
E_n = E_0  - \frac{m \vc^2}{2} \sum_{k=1}^n \sum_{j=1}^n \sigma_{j-k}\cos{\omega(t_j-t_k)}.
\end{equation}
If $E_n>0$, the waves escape to infinity with escape velocity $v_{\rm out} = \pm \sqrt{2 E_n/m}$ and $t_{n+1}$ undefined in~\eqref{eq:dt}. 
A solution is an $n$-bounce resonance if $E_n = E_0$, i.e.\ if 
\begin{multline*}
0 = \sum_{j=1}^n \sum_{k=1}^n \sigma_{j-k}\cos{\omega(t_k-t_j)} \\
= \Re\left( \sum_{j=1}^n \sum_{k=1}^n \sigma_{j-k}e^{i\omega(t_k-t_j)} \right)
= Q_n Q_n^*,
 \end{multline*} i.\ e.\ $Q_n=0$
 where  $Q_n = \sum_{j=1}^{n} \sigma_j e^{i \omega t_j}$ .

 For an exact $n$-bounce resonant solution, $E_{n-j} = E_j$.  Letting $\theta_j = \omega(t_{j+1}-t_j)$,  this implies $\theta_{n-j}=\theta_j$ which yields the following conditions:
\begin{align*}
&\text{\textbf{If}} \, n=2m: && \sum_{j=1}^m (-1)^{j+1}\sin{\bigl ( \sum_{k=j}^m \theta_k-\frac{\theta_m}{2} \bigr)} &=0;\\
&\text{\textbf{If}} \, n=2m+1: && \sum_{j=1}^m (-1)^{j+1}\cos{\bigl (\sum_{k=j}^m \theta_k\bigr) }&=0.
\end{align*}
For model~\eqref{eq:sg_defs}, we find  for 2- and 3-bounce resonances:
\begin{equation}
v_n^{(2)} = \sqrt{ \vc^2 - \frac{\omega^2}{4n^2}}
\text{ and }
v_{n\pm}^{(3)} = \sqrt{ \vc^2 - \frac{\omega^2}{4(n\pm \frac{1}{6})^2}},
\label{eq:vn}
\end{equation}
while for $n\ge 4$ these formulas must be solved numerically, in conjunction with~\eqref{eq:dt}.  

In models~\eqref{eq:sg_defs} and~\eqref{eq:phi4_model}, ${\cal R}=o(1)$ in equation~\eqref{eq:dt}, whereas  ${\cal R}$, calculated numerically, is $O(1)$ for the model studied in~\cite{GooHab:05}. Thus~\eqref{eq:vn} must be modified to include nonzero $\cal R$.

Most initial conditions do not lead to exact $n$-bounce resonances and we approximate the dynamics by iterating the map defined by~\eqref{eq:dt} and~\eqref{eq:DeltaE_general}.  The equivalent to figure~\ref{fig:2bounce}d may be created by varying $E_{\rm in} = m v_{\rm in}^2/2$, and iterating until $E_n=E_{\rm out} = m v_{\rm out}^2/2 >0$. In addition, the graph is color-coded by $n$, the number of bounces preceding escape.  In figures~\ref{fig:sgmap}c and~\ref{fig:sgmap}d we compare solutions of this map to solutions of the sine-Gordon model~\eqref{eq:sg_defs}  and find impressive  agreement for $\epsilon = 0.25$.  The resonant initial velocities given by equation~\eqref{eq:vn} are  marked in figure~\ref{fig:sgmap}c.
In~\cite{GooHab:05} we show, for the map describing~\eqref{eq:phi4}, clusters  of $(n+1)$-bounce windows accumulate at the edges of each $n$-bounce window, repeated at diminishing scales, with an intricacy that would be difficult to achieve from ODE initial-value simulations.
For a wider view of fractal behavior, figure~\ref{fig:sgmap}e shows, using false-color, $n$, the number of bounces or interactions before the kink escapes, as a function of both $\epsilon$ and the initial velocity. The black curve gives $\vc(\epsilon)$, and the fractal structure to its left shows how the windows appear and disappear as $\epsilon$ is varied.

In summary, we have explained the intricate chaotic dynamics arising in the interactions of solitary waves by the dynamics of simple iterated separatrix maps.  Our method applies to many such systems studied over the past 25 years.  We have shown that the 2-bounce resonance is the simplest manifestation of this chaotic scattering process.  

\emph{Acknowledgements:} R.\ G. was supported by NSF  DMS-0204881 and DMS-0506495.  We thank Michel Peyrard and Michael Weinstein for their reading and helpful suggestions.  We dedicate this to Alwyn Scott, who provided valuable suggestions and encouragement.

\end{document}